\documentclass[conference]{IEEEtran}

\usepackage{amssymb}
\usepackage[cmex10]{amsmath}
\ifCLASSOPTIONcaptionsoff
  \usepackage[nomarkers]{endfloat}
 \let\MYoriglatexcaption\caption
 \renewcommand{\caption}[2][\relax]{\MYoriglatexcaption[#2]{#2}}
\fi

\usepackage{tikz}
\usetikzlibrary{positioning,arrows,shapes,decorations.pathmorphing}

\begin{document}

\title{ An Integer Programming Approach to UEP Coding \\ for Multiuser Broadcast Channels}
\author{\IEEEauthorblockN{Wook Jung} \\
\IEEEauthorblockA{The Volgenau School of Engineering\\
George Mason University, Fairfax, VA. 22030 \\
Email: wjung@gmu.edu}\\
\and
\IEEEauthorblockN{Shih-Chun Chang}\\
\IEEEauthorblockA{Department of Electrical and Computer Engineering\\
George Mason University, Fairfax, VA. 22030 \\
Email: schang@gmu.edu}}

\maketitle
\begin{abstract}
In this paper, an integer programming approach is introduced to construct Unequal Error Protection (UEP) codes for multiuser broadcast channels. We show that the optimal codes can be constructed that satisfy the integer programming bound. Based on the bound, we compute asymptotic code rate and perform throughput analysis for the degraded broadcast channel.
\end{abstract}

\section{Introduction}\label{Introduction}

In this paper, we present an integer programming approach to construct unequal error protection (UEP) codes for multiuser broadcast channels. Information theoretic features of the broadcast channel have been studied in \cite{Cover1972,Cover1998,Bergmans1973} and the designs of UEP codes have been well investigated in \cite{Masnick1967,DUNNING1978,BOYARINOV1981,VANGILS1983,VANGILS1984,Lin1990}. The UEP coding scheme corrects unequal channel errors and recovers messages for all users effectively over the broadcast communication system.  

\begin{figure}[!t]
\centering
\begin{tikzpicture}[
	cnode/.style={rectangle,font=\footnotesize},
	rnode/.style={circle,draw,font=\footnotesize\sffamily},
	bnode/.style={rectangle,draw,densely dotted,minimum height=40mm,font=\footnotesize\sffamily},	
	tbox/.style={rectangle,font=\footnotesize\sffamily}, line/.style={->,draw,>=stealth',shorten >=1pt}]
	
	\node (chd) [rectangle,minimum width=20mm,minimum height=36mm] {};
	\node (ch)  [bnode] {\begin{tabular}{c}Broadcast \\ Channel \end{tabular}};
	\node (e) [rnode,left=of ch] {$E$};	
	\node (s) [cnode,left=of e] {$\bullet$};
	\node (m) [cnode,above=of s,xshift=10mm,yshift=-5mm] {$\mathbf{m} = \left( m_1, m_2, \ldots, m_k \right)$};	
		
	\node (d2) [rnode,right=of ch] {$D_2$};	
	\node (d1) [rnode,above=of d2] {$D_1$}; 
	\node (vd) [tbox,below=of d2,yshift=10mm] {$\vdots$};	
	\node (dk) [rnode,below=of d2] {$D_k$};	
	
	\node (r2) [cnode,right=of d2] {$\hat{m}_2$};	
	\node (r1) [cnode,right=of d1] {$\hat{m}_1$}; 
	\node (vr) [tbox,right=of vd] {$\vdots$};	
	\node (rk) [cnode,right=of dk] {$\hat{m}_k$};		
		
	\path (s) edge [line] (e);
	\path (e) edge [line,above,sloped] node (c) {$\mathbf{c}$} (ch);
	\path (chd.north east) edge [line,above,sloped] node (1r) {$\mathbf{r}_1$} (d1);
	\path (chd) edge [line,above,sloped] node (2r) {$\mathbf{r}_2$} (d2);
	\path (chd.south east) edge [line,above,sloped] node (Lr) {$\mathbf{r}_k$} (dk);	
	\path (d1) edge [line,above,sloped] (r1);
	\path (d2) edge [line,above,sloped] (r2);
	\path (dk) edge [line,above,sloped] (rk);		
	
\end{tikzpicture}
\caption{Broadcast communication system}
\label{Model}
\end{figure}
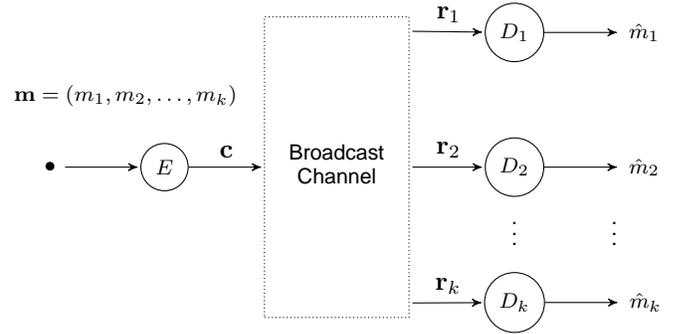

The multiuser broadcast communication system is depicted in Fig. \ref{Model}. Due to the unequal error characteristics of the broadcast channel, each receiver may have different level of error protection requirement denoted by 
\setlength{\arraycolsep}{0.0em}\begin{equation}\label{Tprofile}
	\mathbf{t} = \left( t_1, t_2,  \ldots, t_k \right) \quad\mbox{for}\quad 1\leq i \leq k
\end{equation}\setlength{\arraycolsep}{5pt}
\noindent where $t_i$ is the error protection requirement for a receiver $i$.

Let $C$ be a ($n$, $k$) UEP code with a generator matrix $\mathbf{G}$; 
\setlength{\arraycolsep}{0.0em}\begin{equation*}\label{Cdef}
	C = \left\lbrace \mathbf{m} \mathbf{G} \;|\; m_i \in \lbrace 0, 1\rbrace, \quad 1\leq i\leq k \right\rbrace.
\end{equation*}\setlength{\arraycolsep}{5pt}
Then a separation vector $\mathbf{s} = \left( s_1, s_2 , \ldots, s_k \right)$ of the UEP code $C$ is defined in \cite{DUNNING1978} as follows 
\setlength{\arraycolsep}{0.0em}\begin{equation}\label{Svector}
	s_i = \min \left\lbrace w_{H}\left( \mathbf{m}\mathbf{G}\right) \;\vert\; m_i = 1 \right\rbrace, \quad 1\leq i\leq k
\end{equation}\setlength{\arraycolsep}{5pt} 
\noindent where $w_{H}(\cdot)$ denotes the Hamming weight. Given 
\setlength{\arraycolsep}{0.0em}\begin{equation}\label{Sprofile}
	 s_i \geq 2t_i + 1  \quad\mbox{for}\quad 1\leq i\leq k,
\end{equation}\setlength{\arraycolsep}{5pt}
\noindent the receiver $i$ can correct up to $t_i$ errors and can recover the message $\hat{m}_i$. Given a specific separation vector, we design UPE codes that satisfy (\ref{Svector}) based on an integer programming approach.

In the following sections, we introduce an integer programming model to construct UEP codes and derive the performance bounds to show the efficiency of the codes. Based on the integer programming bound, we compute asymptotic code rate and perform throughput analysis for the degraded broadcast channel.  
  
\section{UEP codes for broadcast channels}\label{UEPcode}

In \cite{Masnick1967}, Masnick and Wolf have presented a UEP code construction method. However, it has been claimed by the authors that the optimal code construction beyond $k=5$ is extremely difficult. It motivates us to develop an innovative approach based on integer programming to construct optimal UEP codes beyond $k=5$.

\subsection{Integer programming approach}\label{IP}

We begin our approach by specifying the separation vector $\mathbf{s} = \left( s_1, s_2, \ldots, s_k \right)$ satisfies, 
\setlength{\arraycolsep}{0.0em}\begin{equation}\label{SVrequirement}
	s_i \leq s_j \quad\mbox{for}\quad 1\leq i < j \leq k.
\end{equation}\setlength{\arraycolsep}{5pt}
Let $\mathbf{A}_a$ be a $k\times (2^k-1)$ matrix consisting of all nonzero binary $k$-tuples as columns in increasing order. 
\setlength{\arraycolsep}{0.0em}\begin{equation}\label{AaDef}
	\mathbf{A}_a = \left( \begin{array}{c@{\;}c@{\;}c@{\;}c@{\;}c@{\;}c@{\;}c@{\;}c@{\;}c}
	0 & 0 & 0 & \cdots & 0 & 1 & \cdots & 1 & 1\\
	0 & 0 & 0 & \cdots & 1 & 0 & \cdots & 1 & 1\\	
	\vdots & \vdots & \vdots & \ddots & \vdots & \vdots & \ddots & \vdots & \vdots\\
	0 & 1 & 1 & \cdots & 1 & 0 & \cdots & 1 & 1\\
	1 & 0 & 1 & \cdots & 1 & 0 & \cdots & 0 & 1	
	\end{array}\right) = \left( \begin{array}{c} \mathbf{a}_1 \\ \mathbf{a}_2 \\ \vdots \\ \mathbf{a}_{k-1} \\ \mathbf{a}_k\end{array}\right)
\end{equation}\setlength{\arraycolsep}{5pt}
\noindent where $\mathbf{a}_i = \left( a_{i,1}, a_{i,2}, \ldots, a_{i,2^k-1} \right)$ is a row vector. Let $\mathbf{G}$ be a generator matrix of UEP code written as
\setlength{\arraycolsep}{0.0em}\begin{equation}\label{GDef}
	\mathbf{G} = \left( \begin{array}{c} \mathbf{g}_1 \\ \mathbf{g}_2 \\ \vdots \\ \mathbf{g}_k\end{array}\right),
\end{equation}\setlength{\arraycolsep}{5pt}
\noindent then each $\mathbf{g}_i$ for $1\leq i\leq k$ can be represented as
\setlength{\arraycolsep}{0.0em}\begin{equation}\label{Grow}
	\mathbf{g}_i = \Bigl( \underbrace{a_{i,1}, \ldots, a_{i,1}}_{x_1}, \underbrace{a_{i,2}, \ldots, a_{i,2}}_{x_2}, \cdots, \underbrace{a_{i,2^k-1}, \ldots, a_{i,2^k-1}}_{x_{2^k-1}} \Bigr)
\end{equation}\setlength{\arraycolsep}{5pt}
\noindent where $\sum_{j=1}^{2^k-1} x_i$ is the code length $n$.

It follows from (\ref{Svector}) that, for a given non-decreasing separation vector $\mathbf{s} = \left( s_1, s_2, \ldots, s_k \right)$,  
\setlength{\arraycolsep}{0.0em}\begin{subequations}\label{GrowSi} \begin{align}
	& w_H \left( \mathbf{g}_1 \right) \geq s_1,  \\
	& w_H \Bigl( \mathbf{g}_i + \sum_{j=1}^{i-1} m_j \mathbf{g}_j \Bigr) \geq s_i & \mbox{for} &\quad 2\leq i\leq k
\end{align}\end{subequations}\setlength{\arraycolsep}{5pt}
\noindent where $ m_j \in \lbrace 0,1 \rbrace$. From (\ref{Grow}), 
\setlength{\arraycolsep}{0.0em}\begin{equation}\label{GrowWeight}
	w_H \left( \mathbf{g}_i \right) = \sum_{j=1}^{2^k-1} a_{i,j} x_j = \mathbf{a}_i \mathbf{x}^\top \quad\mbox{for}\quad 1\leq i\leq k
\end{equation}\setlength{\arraycolsep}{5pt}
\noindent where $\mathbf{x} = \left( x_1, x_2, \ldots, x_{2^k-1} \right)$. Consequently, 
\setlength{\arraycolsep}{0.0em}\begin{align}\label{GrowWeight2}
	w_H \Bigl( \mathbf{g}_i + \sum_{j=1}^{i-1} m_j \mathbf{g}_j \Bigr) &= \left( \mathbf{a}_i + \sum_{j=1}^{i-1} m_j \mathbf{a}_j \right) \mathbf{x}^\top.
\end{align}\setlength{\arraycolsep}{5pt}
From (\ref{GrowSi}), (\ref{GrowWeight}), and (\ref{GrowWeight2}), let $\mathbf{A}_i$ contains all $2^{i-1}$ rows of 
\setlength{\arraycolsep}{0.0em}\begin{subequations}\label{AiRows} \begin{align}
	& \mathbf{a}_i & \mbox{for} & \quad i = 1 \\
	& \mathbf{a}_i + \sum_{j=1}^{i-1} m_j \mathbf{a}_j & \mbox{for} &\quad 2\leq i\leq k,
\end{align}\end{subequations}\setlength{\arraycolsep}{5pt}
then, we have the following inequalities
\setlength{\arraycolsep}{0.0em}\begin{equation}\label{AiINEQ}
	\mathbf{A}_i \mathbf{x}^\top \geq \mathbf{b}_i^\top \quad\mbox{for}\quad 1\leq i\leq k
\end{equation}\setlength{\arraycolsep}{5pt}
\noindent where $\mathbf{A}_i$ is a $2^{i-1}\times (2^k-1)$ matrix and $\mathbf{b}_i = \left( s_i, s_i, \ldots, s_i\right)$ is a row vector of length $2^{i-1}$.
Based on (\ref{AiINEQ}), we can formulate UEP code construction for a given non-decreasing separation vector $\mathbf{s} = \left( s_1, s_2, \ldots, s_k \right)$ as an integer programming model.

\vspace{1em}
\framebox[3.2in][c]{\parbox{2.9in}{\vspace{2mm} \underline{UEP code construction with integer programming}  \vspace{3mm} \\ \begin{IEEEeqnarraybox}{r.C.}
	\mbox{Minimize} \quad & n_{\mathbf{s}} = x_1 + x_2 + \cdots + x_{2^k-1} \\
	\noalign{ subject to} & \mathbf{A} \mathbf{x}^\top \geq \mathbf{b}^\top \\
	\noalign{ where} & \mathbf{A} = \left( \begin{array}{c} \mathbf{A}_1 \\ \mathbf{A}_2 \\ \vdots \\ 		\mathbf{A}_k \end{array}\right) , \quad \mathbf{b}^\top = \left( \begin{array}{c} \mathbf{b}_1^\top \\ \mathbf{b}_2^\top \\ \vdots \\ \mathbf{b}_k^\top \end{array}\right)
\end{IEEEeqnarraybox}\vspace{3mm}}}
\vspace{1em}

\newtheorem{exam}{\textbf{Example}}
\begin{exam}\label{Example1} For $k=3$ and $\mathbf{s} = \left( 3, 5, 7\right)$,
\setlength{\arraycolsep}{0.0em}\begin{equation*}\label{AaNum}
	\mathbf{A}_a = \left( \begin{array}{ccccccc} 
	0 & 0 & 0 & 1 & 1 & 1 & 1 \\
	0 & 1 & 1 & 0 & 0 & 1 & 1 \\
	1 & 0 & 1 & 0 & 1 & 0 & 1 \\
	\end{array}\right) = \left( \begin{array}{c} \mathbf{a}_1 \\ \mathbf{a}_2 \\ \mathbf{a}_3 \end{array}\right),
\end{equation*}\setlength{\arraycolsep}{5pt}
\noindent and
\setlength{\arraycolsep}{0.0em}\begin{equation*}\label{GNum}
{\mathbf{G} \atop } {= \atop } {\left( \begin{array}{ccc|ccc|c|ccc}
	0 \;&\; \cdots \;&\; 0 \;&\; 0 \;&\; \cdots \;&\; 0 \;&\;        \;&\; 1 \;&\; \cdots \;&\; 1 \\
	0 \;&\; \cdots \;&\; 0 \;&\; 1 \;&\; \cdots \;&\; 1 \;&\; \cdots \;&\; 1 \;&\; \cdots \;&\; 1 \\ 
	1 \;&\; \cdots \;&\; 1 \;&\; 0 \;&\; \cdots \;&\; 0 \;&\;        \;&\; 1 \;&\; \cdots \;&\; 1 \end{array} \right) \atop \quad\leftarrow \hfill x_1 \hfill \rightarrow \leftarrow \hfill x_2 \hfill \rightarrow \cdots \leftarrow \hfill x_7 \hfill \rightarrow\quad} {= \atop} { \left( \begin{array}{c} \mathbf{g}_1 \\ \mathbf{g}_2 \\ \mathbf{g}_3\end{array}\right) \atop }.
\end{equation*}\setlength{\arraycolsep}{5pt}
Then, 
\setlength{\arraycolsep}{0.0em}\begin{IEEEeqnarray*}{ccccl}\label{AiNum}
	\mathbf{A}_1 &=& \left( \begin{array}{ccccccc} 
	0 & 0 & 0 & 1 & 1 & 1 & 1  \end{array}\right)  &=& \left( \mathbf{a}_1 \right),  \\[1em]
	\mathbf{A}_2 &=& \left( \begin{array}{ccccccc} 
	0 & 1 & 1 & 0 & 0 & 1 & 1 \\
	0 & 1 & 1 & 1 & 1 & 0 & 0 \end{array}\right)  &=& \left( \begin{array}{l} \mathbf{a}_2 \\ \mathbf{a}_2 + \mathbf{a}_1\end{array}\right), \\[1em]
	\mathbf{A}_3 &=& \left( \begin{array}{ccccccc} 
	1 & 0 & 1 & 0 & 1 & 0 & 1 \\
	1 & 0 & 1 & 1 & 0 & 1 & 0 \\
	1 & 1 & 0 & 0 & 1 & 1 & 0 \\
	1 & 1 & 0 & 1 & 0 & 0 & 1 \end{array}\right)  &=& \left( \begin{array}{l} \mathbf{a}_3 \\ \mathbf{a}_3 + \mathbf{a}_1 \\ \mathbf{a}_3 + \mathbf{a}_2 \\ \mathbf{a}_3 + \mathbf{a}_1 + \mathbf{a}_2\end{array}\right). 
\end{IEEEeqnarray*}\setlength{\arraycolsep}{5pt}
Therefore, the UEP code construction is to find $\mathbf{x}$ that minimizes the sum $x_1 + x_2 + \cdots + x_7$ satisfying 
\setlength{\arraycolsep}{0.0em}\begin{IEEEeqnarray*}{c;c;c}\label{Xnum}
 	x_4 + x_5 + x_6 + x_7 & \geq& 3 \\
    x_2 + x_3 + x_6 + x_7 & \geq& 5 \\
    x_2 + x_3 + x_4 + x_5 & \geq& 5 \\
    x_1 + x_3 + x_5 + x_7 & \geq& 7 \\
    x_1 + x_3 + x_4 + x_6 & \geq& 7 \\
    x_1 + x_2 + x_5 + x_6 & \geq& 7 \\
    x_1 + x_2 + x_4 + x_7 & \geq& 7 .
\end{IEEEeqnarray*}\setlength{\arraycolsep}{5pt}
The optimal solution for this example shall be given in later section.
\end{exam}

\subsection{Integer programming bound}\label{Bounds}

In \cite{VANGILS1983}, van Gils has derived a lower bound of a UEP code for a non-increasing separation vector $\mathbf{s}' = \left( s'_1, s'_2, \ldots, s'_k \right)$ as 

\setlength{\arraycolsep}{0.0em}\begin{IEEEeqnarray}{rCl}\label{VGorigin}
	n_{\mathbf{s}'} &\geq& \sum_{i=1}^{k} \;\left\lceil \frac{s'_i}{2^{i-1}} \right\rceil.
\end{IEEEeqnarray}\setlength{\arraycolsep}{5pt}

We begin by introducing the bound of (\ref{VGorigin}) for a purpose to give the following theorem.
\vspace{0.5em}
\newtheorem{thm}{\textbf{Theorem}}
\begin{thm}[Integer programming bound]\label{IPbound}
For a given $k$, let $\mathbf{s} = \left( s_1, s_2, \ldots, s_k \right)$ be a non-decreasing separation vector, then the integer programming bound is given by
\setlength{\arraycolsep}{0.0em}\begin{equation}\label{UEPbound}
	n_{\mathbf{s}} \geq \sum_{i=1}^{k} \left\lceil \frac{s_i}{2^{k-i}} \right\rceil. 
\end{equation}\setlength{\arraycolsep}{5pt}
\end{thm}

\begin{IEEEproof}
For the integer programming problem, we formulate $2^k-1$ inequalities in matrix form as
\[\mathbf{A} \mathbf{x}^\top \geq \mathbf{b}^\top.\]
Since the rows of $\mathbf{A}$ consist of $2^k-1$ linear combinations of $\mathbf{a}_i$ for $1\leq i\leq k$, each column of $\mathbf{A}$ has $2^{k-1}$ ones \cite[p.97]{LinBook}. Therefore, the sum of $2^k-1$ inequalities is
\setlength{\arraycolsep}{0.0em}\begin{equation*}
	2^{k-1} \left( x_1 + x_2 + \cdots + x_{2^k-1} \right) \geq \sum_{i=1}^{k} 2^{i-1} s_i,
\end{equation*}\setlength{\arraycolsep}{5pt}
\noindent hence,
\setlength{\arraycolsep}{0.0em}\begin{equation}\label{IPboundOrigin}
 n_\mathbf{s} \;\geq\; \sum_{i=1}^{k} \frac{s_i}{2^{k-i}}.
\end{equation}\setlength{\arraycolsep}{5pt}
Since the separation vector $\mathbf{s} = \left( s_1, s_2, \ldots, s_k \right)$ is in non-decreasing order, set $s_i' = s_{k-i+1}$ and then, transform the bound of (\ref{VGorigin}) into  
\setlength{\arraycolsep}{0.0em}\begin{equation}\label{VGbound}
	n_{\mathbf{s}'} \geq \sum_{i=1}^{k} \;\left\lceil \frac{s'_i}{2^{i-1}} \right\rceil = \sum_{i=1}^{k} \left\lceil \frac{s_{k-i+1}}{2^{i-1}} \right\rceil = \sum_{i=1}^{k} \left\lceil \frac{s_{i}}{2^{k-i}} \right\rceil.
\end{equation}\setlength{\arraycolsep}{5pt}
For every $i$ from $1$ to $k$,
\setlength{\arraycolsep}{0.0em}\begin{equation}\label{BoundComp}
	\left\lceil \frac{s_{i}}{2^{k-i}} \right\rceil \;\geq\; \frac{s_{i}}{2^{k-i}}.
\end{equation}\setlength{\arraycolsep}{5pt}
Therefore, the actual lower bound of the integer programming is exactly the same as the bound of (\ref{VGbound}). Consequently
\setlength{\arraycolsep}{0.0em}\begin{equation*}
	n_{\mathbf{s}} \;\geq\; \sum_{i=1}^{k} \left\lceil \frac{s_{i}}{2^{k-i}} \right\rceil. 
\end{equation*}\setlength{\arraycolsep}{5pt}
\end{IEEEproof}

In the following, we also include an upper bound on the length $n_{\mathbf{s}}$ derived by Masnick and Wolf in \cite{Masnick1967}, 
\setlength{\arraycolsep}{0.0em}\begin{equation}\label{Wbound}
	n_{\mathbf{s}} \leq k + r_{U} 
\end{equation} \setlength{\arraycolsep}{5pt}
\noindent and $r_{U}$ is the smallest number of check bits $r$ such that $2^r > \gamma$ where 
\setlength{\arraycolsep}{0.0em}\begin{equation}\label{Wolf}
	\gamma = \sum_{i=0}^{2t_k-1} \binom{n-1}{i} - \sum_{i=2}^{k} \binom{n-1-T_i}{2t_i-2} - \sum_{i=2}^{k} \binom{n-1-T_i}{2t_i-1}
	\end{equation}\setlength{\arraycolsep}{5pt}
\noindent and $T_i$ is the number of message bits that are protected against $i$-bit or more errors. Numerical results for the lower and upper bounds are shown in Table \ref{Rtable}.

\subsection{Integer programming results}\label{IPresults}

\begin{figure}[!t]
\centering
\includegraphics[width=3in]{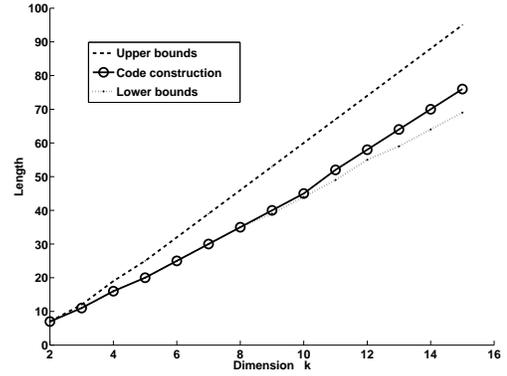}
\caption{Integer programming results with bounds.}
\label{FIGbounds}
\end{figure}

In Table \ref{Rtable}, we present the numerical results of integer programming by using IBM ILOG CPLEX v$12.5.1$ \cite{CPLEX} for $\mathbf{s} = \left( 3, 5,\ldots, 2k+1 \right)$. From $k=2$ to $8$, $n_{\mathbf{s}}$ are exactly the same as the lower bound, therefore the corresponding UEP codes are optimal. The optimal results are listed in Table \ref{IPtable}. For $9\leq k\leq 15$, instead of finding optimal solutions, we limit our search for sub-optimal solutions due to time constraint. The results and both lower and upper bounds are plotted in Fig. \ref{FIGbounds}.

\begin{table}[!t]
\renewcommand{\arraystretch}{1.3}
\caption{Numerical results for $\mathbf{s} = \left( 3, 5, \ldots, 2k+1 \right)$} 
\label{Rtable}
\centering
	\begin{tabular}{c|cccc}
	$k$  & \shortstack{Lower bounds \\ (\ref{UEPbound})} & \shortstack{Integer programming \\ results, $n_{\mathbf{s}}$} & \shortstack{Upper bounds \\ (\ref{Wbound})} \\\hline\hline
	2  	& 7 		& 7	   & 7  	 \\
	3  	& 11 		& 11   & 12   	 \\
	4  	& 16 		& 16   & 19 	 \\
	5  	& 20 		& 20   & 25 	 \\
	6  	& 25 		& 25   & 32 	 \\
	7  	& 30 		& 30   & 39 	 \\
	8  	& 35 		& 35   & 46 	 \\ 
	9  	& 39 		& $\langle 40\rangle$   & 53 	 \\ 
	10 	& 44 		& $\langle 45\rangle$   & 60 	 \\ 
	11 	& 49 		& $\langle 52\rangle$   & 67 	 \\ 
	12 	& 55 		& $\langle 58\rangle$   & 74 	 \\ 
	13 	& 59 		& $\langle 64\rangle$   & 81 	 \\ 
	14 	& 64 		& $\langle 70\rangle$   & 88 	 \\ 
	15 	& 69 		& $\langle 76\rangle$   & 95 	 \\ \hline
	\end{tabular}\\[1ex]
	Note: $\langle \cdot \rangle$ indicates sub-optimal length.	
\end{table}

\vspace{0.5em}
\begin{exam} 
The optimal solution obtained from the integer programming for $\mathbf{s}=\left( 3, 5, 7\right)$ is
\setlength{\arraycolsep}{0.0em}\begin{equation*}
	n_{\mathbf{s}} = \sum_{i=1}^{7} x_i = 11,
\end{equation*}\setlength{\arraycolsep}{5pt}
and one of the optimal solutions occurs at
\setlength{\arraycolsep}{0.0em}\begin{equation*}
	\begin{array}{c@{\;}c@{\;}c@{\;}c@{\;}c@{\;}c@{\;}c}
	x_1 & x_2 & x_3 & x_4 & x_5 & x_6 & x_7  \\ \hline
	3 & 2 & 2 & 1 & 1 & 1 & 1 \end{array}.
\end{equation*}\setlength{\arraycolsep}{5pt}
It means that $i$-th column of $\mathbf{A}_a$ appears $x_i$ times in the optimal generator matrix. Therefore, 
\setlength{\arraycolsep}{0.0em}\begin{equation*}
	\mathbf{G} = \left( \begin{array}{c@{\;}c@{\;}c@{\;}c@{\;}c@{\;}c@{\;}c@{\;}c@{\;}c@{\;}c@{\;}c} 
	0 & 0 & 0 & 0 & 0 & 0 & 0 & 1 & 1 & 1 & 1 \\
	0 & 0 & 0 & 1 & 1 & 1 & 1 & 0 & 0 & 1 & 1 \\
	1 & 1 & 1 & 0 & 0 & 1 & 1 & 0 & 1 & 0 & 1 
	\end{array}\right).
\end{equation*}\setlength{\arraycolsep}{5pt}
Let $\hat{\mathbf{s}} = \left( \hat{s}_1, \hat{s}_2, \ldots, \hat{s}_k\right)$ denote the separation vector obtained from the corresponding $\mathbf{G}$. If
\setlength{\arraycolsep}{0.0em}\begin{equation*}
	\hat{s}_i \geq s_i \quad\mbox{for}\quad 1\leq i\leq k,
\end{equation*}\setlength{\arraycolsep}{5pt}
then $\mathbf{G}$ satisfies (\ref{Svector}). In this example,
\setlength{\arraycolsep}{0.0em}\begin{equation*}
	\hat{\mathbf{s}} = \left( 4, 6, 7 \right) \quad\geq\quad \mathbf{s} = \left( 3, 5, 7 \right).
\end{equation*}\setlength{\arraycolsep}{5pt}
\end{exam}

There are many combinations of $x_i$ for $1\leq i \leq 2^k-1$ that are summed to the optimal value of $n_{\mathbf{s}}$ (e.g., there are $3$ solutions of $x_i$ to $n_{\mathbf{s}} = 7$ when $k=2$, $7$ solutions of $x_i$ to $n_{\mathbf{s}} = 11$ when $k=3$, $2063$ solutions of $x_i$ to $n_{\mathbf{s}} = 16$ when $k=4$, and so on).
\begin{table*}[!t]
\renewcommand{\arraystretch}{1.3}
\caption{Optimal results of integer programming}
\label{IPtable}
\centering
	\setlength{\arraycolsep}{0.0em}\begin{tabular}{llll}
	$k$ & $n_{\mathbf{s}}$ 	& non-zero $x_i$ for $1\leq i\leq 2^k-1$ 	& $\hat{\mathbf{s}}$   \\ \hline\hline
	2  	& 7	& $\begin{array}{*{3}{c}} x_{1} & x_{2} & x_{3} \\ \hline 4 & 2 & 1 \end{array}$	& $\left( 3, 5 \right)$  	 \\
	3  	& 11	& $\begin{array}{*{7}{c}} x_{1} & x_{2} & x_{3} & x_{4} & x_{5} & x_{6} & x_{7} \\ \hline 3 & 2 & 2 & 1 & 1 & 1 & 1 \end{array}$	& $\left( 4, 6, 7 \right)$   	 \\
	4  	& 16	& $\begin{array}{*{15}{c}}
x_{1} & x_{2} & x_{3} & x_{4} & x_{5} & x_{6} & x_{7} & x_{8} & x_{9} & x_{10} & x_{11} & x_{12} & x_{13} & x_{14} & x_{15} \\ \hline 
2 & 1 & 1 & 1 & 1 & 1 & 1 & 1 & 1 & 1 & 1 & 1 & 1 & 1 & 1 \end{array}$	& $\left( 8, 8, 8, 9\right)$ 	 \\
	5  	& 20 	& $\begin{array}{*{19}{c}}
x_{1} & x_{2} & x_{3} & x_{4} & x_{5} & x_{6} & x_{7} & x_{8} & x_{9} & x_{10} & x_{11} & x_{12} & x_{13} & x_{14} & x_{15} & x_{16} & x_{17} & x_{18} & x_{19} \\ \hline 
2 & 1 & 1 & 1 & 1 & 1 & 1 & 1 & 1 & 1 & 1 & 1 & 1 & 1 & 1 & 1 & 1 & 1 & 1 \end{array} $	& $\left( 4, 8, 8, 10, 11\right)$  	 \\
	6  	& 25 & $\begin{array}{*{20}{c}}
x_{1} & x_{2} & x_{3} & x_{4} & x_{5} & x_{6} & x_{7} & x_{8} & x_{9} & x_{10} & x_{11} & x_{12} & x_{13} & x_{14} & x_{15} & x_{16} & x_{17} & x_{18} & x_{19} & x_{20} \\ \hline 
1 & 1 & 1 & 1 & 1 & 1 & 1 & 1 & 1 & 1 & 1 & 1 & 1 & 1 & 1 & 1 & 1 & 1 & 1 & 1 \\ 
x_{21} & x_{32} & x_{33} & x_{34} & x_{35} \\ \cline{1-5} 
1 & 1 & 1 & 1 & 1 \end{array}$ & $\left( 4, 6, 8, 10, 12, 13\right)$ \\
	7  	& 30 & $\begin{array}{*{20}{c}}
x_{1} & x_{2} & x_{3} & x_{4} & x_{5} & x_{6} & x_{7} & x_{8} & x_{9} & x_{10} & x_{11} & x_{12} & x_{13} & x_{14} & x_{15} & x_{16} & x_{17} & x_{18} & x_{19} & x_{20} \\ \hline 
1 & 1 & 1 & 1 & 1 & 1 & 1 & 1 & 1 & 1 & 1 & 1 & 1 & 1 & 1 & 1 & 1 & 1 & 1 & 1 \\ 
x_{21} & x_{22} & x_{23} & x_{32} & x_{33} & x_{34} & x_{35} & x_{64} & x_{65} & x_{104} \\ \cline{1-10} 
1 & 1 & 1 & 1 & 1 & 1 & 1 & 1 & 1 & 1 \end{array}$ &  $\left( 3, 5, 8, 9, 12, 14, 15\right)$	 \\ 
	8  	& 35 & $\begin{array}{*{20}{c}}
x_{1} & x_{2} & x_{3} & x_{4} & x_{5} & x_{6} & x_{7} & x_{8} & x_{9} & x_{10} & x_{11} & x_{12} & x_{13} & x_{14} & x_{15} & x_{16} & x_{17} & x_{18} & x_{19} & x_{20} \\ \hline 
1 & 1 & 1 & 1 & 1 & 1 & 1 & 1 & 1 & 1 & 1 & 1 & 1 & 1 & 1 & 1 & 1 & 1 & 1 & 1 \\ 
x_{21} & x_{22} & x_{23} & x_{32} & x_{33} & x_{34} & x_{35} & x_{64} & x_{65} & x_{66} & x_{126} & x_{127} & x_{128} & x_{129} & x_{184} \\ \cline{1-15} 
1 & 1 & 1 & 1 & 1 & 1 & 1 & 1 & 1 & 1 & 1 & 1 & 1 & 1 & 1 \end{array}$ & $\left( 3, 5, 7, 11, 11, 13, 15, 17\right)$ 	 \\ \hline
	\end{tabular}\setlength{\arraycolsep}{5pt}
\end{table*}

\subsection{Asymptotic code rates}\label{LBanalysis}

For $\mathbf{s}=\left( 3, 5, \ldots, 2k+1 \right)$, we present the corresponding asymptotic code rates based on the lower bounds as follows.  
\vspace{0.5em}
\begin{thm}\label{AsympRATE}
Let $R_{\mathsf{UEP}}$ be the rate of UEP code whose separation vector is $\mathbf{s} = \left( 3, 5, \ldots, 2k+1\right)$. Then 
\setlength{\arraycolsep}{0.0em}\begin{equation}\label{CRasymp}	
	R_{\mathsf{UEP}} \approx 0.2 \quad\mbox{when}\quad k \gg 1.
\end{equation}\setlength{\arraycolsep}{5pt}
\end{thm}

\begin{figure}[!t]
\centering
\includegraphics[width=3in]{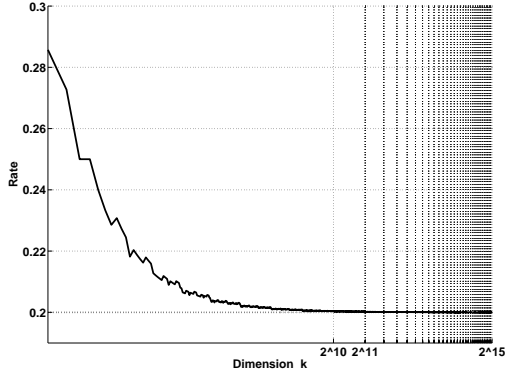}
\caption{UEP code rates for $2\leq k\leq 2^{15}$.}
\label{FIGrate}
\end{figure}

\begin{IEEEproof}
Let $n_{\mathbf{s}}$ satisfy the bound of (\ref{UEPbound}), then
\setlength{\arraycolsep}{0.0em}\begin{equation}
	n_{\mathbf{s}} = \sum_{i=1}^{k} \left\lceil \frac{ 2i + 1}{2^{k-i}} \right\rceil = \sum_{j=0}^{k-1} \left\lceil \frac{ 2(k-j) + 1}{2^{j}} \right\rceil. 
\end{equation}\setlength{\arraycolsep}{5pt}

Let $\eta_j \triangleq  \left\lceil \left( 2k -2j + 1\right) /2^{j} \right\rceil$ and $k = a_x 2^x + a_{x-1} 2^{x-1} + \cdots + a_0$ where $x = \left\lfloor \log_2 k\right\rfloor$, then 
\setlength{\arraycolsep}{0.0em}\begin{equation}
	\eta_j = \left\lbrace \begin{array}{cc}	
	2k + 1 &, \;j=0 \\
	k &, \;j=1 \\	
	a_{x} 2^{x-j+1} + \cdots + a_{j-1} 2^0 +  \delta_j &, 2\leq j\leq x+1\\
	1 &, \;x+2\leq j\leq k-1\end{array}\right.
\end{equation}\setlength{\arraycolsep}{5pt}
\noindent where $\delta_j \in \lbrace 0, 1\rbrace$ \cite[p.47,51]{KnuthBook}.

Let $\Delta_j \triangleq \eta_j - \delta_j$ for $2\leq j\leq x+1$, then 
\setlength{\arraycolsep}{0.0em}\begin{equation}
	\Delta_j = a_{x} 2^{x-j+1} + a_{x-1} 2^{x-j} + \cdots + a_{j-1} 2^0, 
\end{equation}\setlength{\arraycolsep}{5pt}
and
\setlength{\arraycolsep}{0.0em}\begin{IEEEeqnarray}{rCl}
	\sum_{j=2}^{x+1} \Delta_j &=& a_x \left( 2^{x-1} + \cdots + 1 \right) + a_{x-1} \left( 2^{x-2} + \cdots + 1 \right) \IEEEnonumber\\
	&& + \cdots + a_2 \left( 2 + 1 \right) + a_1  \IEEEnonumber \\
	&=& a_x \left( 2^x - 1 \right) + a_{x-1} \left( 2^{x-1} - 1 \right) + \cdots +  a_{0} \left( 1 - 1 \right)\IEEEnonumber\\
	&=& k - \sum_{i=0}^{x} a_i \IEEEyesnumber%
\end{IEEEeqnarray}\setlength{\arraycolsep}{5pt}
Therefore,
\setlength{\arraycolsep}{0.0em}\begin{IEEEeqnarray}{rCl}
	\sum_{j=0}^{1} \eta_j &=& 3k+1, \IEEEyessubnumber\\
	\sum_{j=2}^{x+1} \eta_j &=& k - \sum_{i=0}^{x} a_i + \sum_{j=2}^{x+1} \delta_j, \IEEEyessubnumber\\	
	\noalign{\noindent and\vspace{\jot}}\sum_{j=x+2}^{k-1} \eta_j &=& k-\left\lfloor \log_2 k\right\rfloor-2. \IEEEyessubnumber%
\end{IEEEeqnarray}\setlength{\arraycolsep}{5pt}
Consequently, 
\setlength{\arraycolsep}{0.0em}\begin{equation}
	n_{\mathbf{s}} \;=\; \sum_{j=0}^{k-1} \eta_j \;\approx\; 5k \quad\mbox{when}\quad k\gg 1.
	\end{equation}\setlength{\arraycolsep}{5pt}
Therefore the code rate is converged to $R_{\mathsf{UEP}} \approx \frac{k}{5k} = 0.2$.
\end{IEEEproof}
\vspace{0.5em}
The actual rates of UEP codes based on $R = \frac{k}{n_{\mathbf{s}}}$ are illustrated in Fig. \ref{FIGrate} for $k \leq 2^{15}$, which shows asymptotic convergence to 0.2 as proven in Theorem \ref{AsympRATE}.

\subsection{Throughput of degraded broadcast channels}

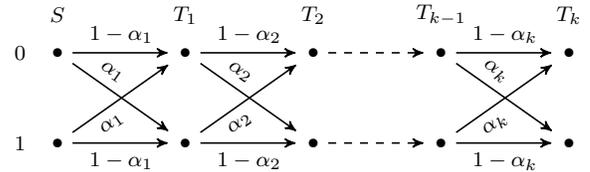
\begin{figure}[!t]
\centering
\begin{tikzpicture}[
	inode/.style={node distance=1.7cm,font=\footnotesize\sffamily},
	tbox/.style={rectangle,font=\footnotesize\sffamily},
	lane/.style={->,>=stealth',shorten >=1pt,semithick,font=\footnotesize\sffamily}]
   	
  	\node	(B1) [inode] {$\bullet$};
  	\node	(B2) [inode,below of=B1,yshift=5mm] {$\bullet$};
  	\node	(Ba) [tbox,left of=B1,xshift=5mm] {$0$};
  	\node	(Bb) [tbox,left of=B2,xshift=5mm] {$1$};
	\node	(B0) [tbox,above of=B1,yshift=-5mm] {$S$};

  	\node	(C1) [inode,right of=B1] {$\bullet$};
  	\node	(C2) [inode,below of=C1,yshift=5mm] {$\bullet$};  	
  	\node	(C0) [tbox,above of=C1,yshift=-5mm] {$T_1$};
  	
  	\node	(D1) [inode,right of=C1] {$\bullet$};
  	\node	(D2) [inode,below of=D1,yshift=5mm] {$\bullet$};
  	\node	(D0) [tbox,above of=D1,yshift=-5mm] {$T_2$};
  	
  	\node	(E1) [inode,right of=D1] {$\bullet$};
  	\node	(E2) [inode,below of=E1,yshift=5mm] {$\bullet$};
  	\node	(E0) [tbox,above of=E1,yshift=-5mm] {$T_{k-1}$};
  	
  	\node	(F1) [inode,right of=E1] {$\bullet$};
  	\node	(F2) [inode,below of=F1,yshift=5mm] {$\bullet$};
	\node	(F0) [tbox,above of=F1,yshift=-5mm] {$T_k$};  	  	
  			
  	\path  	(B1) edge [lane,above] 				node {$1-\alpha_1$} 		(C1)
  				 edge [lane,sloped,above left]	node {$\alpha_1$} 	(C2); 
  	\path  	(B2) edge [lane,sloped,below left] 	node {$\alpha_1$} 	(C1)
  				 edge [lane,below]				node {$1-\alpha_1$} 	(C2); 
  				 			 
  	\path  	(C1) edge [lane,above] 				node {$1-\alpha_2$} 		(D1)
  				 edge [lane,sloped,above left]	node {$\alpha_2$} 	(D2); 
  	\path  	(C2) edge [lane,sloped,below left] 	node {$\alpha_2$} 	(D1)
  				 edge [lane,below]				node {$1-\alpha_2$} 	(D2); 
  				 
  	\path   (D1) edge [lane,dashed,above]	node {} (E1);
  	\path   (D2) edge [lane,dashed,above]	node {} (E2);
  	
  	\path  	(E1) edge [lane,above] 				node {$1-\alpha_k$} 		(F1)
  				 edge [lane,sloped,above left]	node {$\alpha_k$} 	(F2); 
  	\path  	(E2) edge [lane,sloped,below left] 	node {$\alpha_k$} 	(F1)
  				 edge [lane,below]				node {$1-\alpha_k$} 	(F2);   	
\end{tikzpicture}
\caption{Cascaded BSCs for degraded broadcast channel \cite{Bergmans1973}.}
\label{BSCs}
\end{figure}

The proposed broadcast channel model can be considered as a degraded broadcast channel that has $k$ component channels \cite{Bergmans1973}. Let $S$ be a source input and $T_i$ be outputs of the component channels where $S, T_i \in \lbrace 0,1 \rbrace$ for $1\leq i\leq k$, then the error characteristics of the channel can be described with $k$ cascaded binary symmetric channels (BSCs) with parameters $\alpha_i \in [0,1]$ as shown in Fig. \ref{BSCs}.

\begin{figure}[!t]
\centering
\includegraphics[width=3.3in]{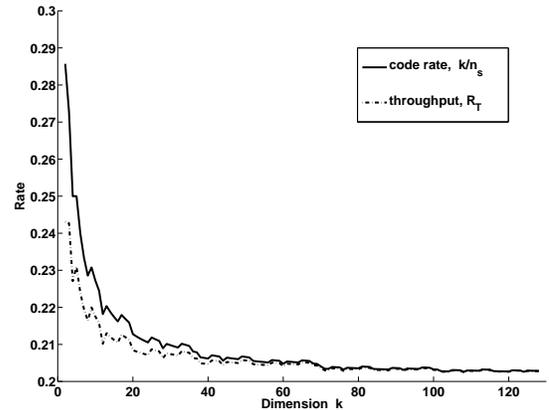}
\caption{Code rate and throughput of the degraded broadcast channel for $2\leq k\leq 2^7$.}
\label{FIGcomb}
\end{figure}

Let $p_i$ denote the bit error probability of the each component channel, then 
\setlength{\arraycolsep}{0.0em}\begin{equation}\label{BERs}	
	p_i = p_{i-1} (1-\alpha_i) + (1-p_{i-1}) \alpha_i, \quad 1\leq i\leq k
\end{equation}\setlength{\arraycolsep}{5pt}
\noindent where $p_0 = 0$ \cite{Bergmans1973}. Let $n_{\mathbf{s}}$ satisfy the bound in (\ref{UEPbound}) for a given separation vector $\mathbf{s} = \left( 3, 5, \ldots, 2k+1 \right)$, then $m_i$ can be successfully delivered to receiver $i$ when the corresponding component channel introduces less than or equal to $i$ errors. Therefore the average rate of successful transmission of a bit to receiver $i$, denoting by $\theta_{i}$, can be given as  
\setlength{\arraycolsep}{0.0em}\begin{equation}\label{Ps}	
	\theta_i = \sum_{j=0}^{i} \binom{n_{\mathbf{s}}}{j} (1-p_i)^{n_{\mathbf{s}}-j} p_i^{j}, \quad 1\leq i\leq k,
\end{equation}\setlength{\arraycolsep}{5pt}
consequently,  
\setlength{\arraycolsep}{0.0em}\begin{equation}\label{Arates}	
	R_i = \frac{\theta_i}{n_{\mathbf{s}}}, \quad 1\leq i\leq k
\end{equation}\setlength{\arraycolsep}{5pt}
where $R_i$ denote the effective transmission rate of each component channel. 
Let $R_T$ be the throughput of the broadcast channel, then 
\setlength{\arraycolsep}{0.0em}\begin{equation}\label{TotalArates}	
	R_T = \sum_{i=1}^{k} R_i \;\leq\; \frac{k}{n_{\mathbf{s}}}.
\end{equation}\setlength{\arraycolsep}{5pt}
The numerical results of the throughput $R_T$, in comparison to the corresponding code rate $\frac{k}{n_{\mathbf{s}}}$, for $2\leq k\leq 2^7$ are illustrated in Fig. \ref{FIGcomb} when $\alpha_i = 1 / n_{\mathbf{s}}$ for $1\leq i\leq k$.  

\section{Conclusion}\label{Conclusion}

In this paper, we have proposed an integer programming approach to UEP coding for multiuser broadcast channels. We have constructed optimal UEP codes by using an integer programming model. Based on the integer programming bound, we have demonstrated the achievable asymptotic code rates in comparison to the throughput of the degraded broadcast channel.

Even though we limit to send $1$ bit to each receiver, the integer programming approach can be extended to send a multi-bit message or a packet to each receiver, we refer this to UEP network coding in a follow-up paper.

\bibliographystyle{IEEEtran}
\bibliography{IEEEabrv,references}
\nocite{*}

\end{document}